
\input harvmac
\noblackbox
\def\Title#1#2{\rightline{#1}\ifx\answ\bigans\nopagenumbers\pageno0\vskip1in
\else\pageno1\vskip.8in\fi \centerline{\titlefont #2}\vskip .5in}

%
%
\def\ajou#1&#2(#3){\ \sl#1\bf#2\rm(19#3)}

\def\hf{{1\over2}}

\font\ticp=cmcsc10

\def\gradbar{\bar\nabla}
\def\chden{\Delta_{\hbox{c}}}
\def\dee{\partial}
\def\deebar{\bar\partial}

\def\lor{M}
\def\why{Y}
\def\act{{\cal L}}
\def\imath{i}
\def\Re{\hbox{Re}}
\def\Im{\hbox{Im}}
\def\R{\Sigma}
\def\pphi{\varphi}
\def\frth{{1\over 4}}
%
%
\lref\Howe{P. S. Howe and G. Papadopoulos, ``$N=2$, $d=2$ Supergeometry",
    \ajou Class. Quantum Grav. &4, (87) 11.}
\lref\CGHS{C.G. Callan, S.B. Giddings, J.A. Harvey, and A. Strominger,
  ``Evanescent Black Holes,"\ajou Phys. Rev. &D45 (92) R1005.}
\lref\RV{For recent reviews see J. A. Harvey and A. Strominger,
  ``Quantum Aspects of Black Holes'' preprint EFI-92-41, hep-th@xxx/9209055,
  to appear in the proceedings of the 1992 TASI Summer School in Boulder,
  Colorado, and S. B. Giddings, ``Toy Models for Black Hole Evaporation''
  preprint UCSBTH-92-36, hep-th@xxx/9209113, to appear in the proceedings
  of the International Workshop of Theoretical Physics, 6th Session,
  June 1992, Erice, Italy.}
\lref\WB{J. Wess and J. Bagger, ``Supersymmetry and Supergravity'',
 2nd edition, Princeton University Press, Princeton (1992).}
\lref\Aln{ Aziz Alnowaiser, ``Supergravity with $N=2$ in two dimensions,''
  \ajou Class. Quantum Grav. &7 (90) 1033.}
\lref\PaSt{Youngchul Park and Andrew Strominger,``Supersymmetry and
  Positive Energy Theorems in Classical and Quantum Two Dimensional
  Dilaton Gravity,''to appear in  Phys. Rev. D, hep-th@xxx/9210017.}
\lref\GiSt{S. Giddings and A. Strominger, ``Quantum Theories of Dilaton
  Gravity'', UCSB-TH-92-28, hep-th@xxx/9207034.}
\lref\Strm{A. Strominger, ``Fadeev-Popov Ghosts and 1 + 1 Dimensional
  Black Hole Evaporation'', UCSB-TH-92-18, hep-th@xxx/9205028.}
\lref\GLO{S. J. Gates Jr., L. Lu, and R. N. Oerter, ``Simplified SU(2)
  Spinning String Superspace Supergravity'', Phys. Lett. {\bf 218B},
  33 (1989).}
\lref\GGRS{S. J. Gates Jr., M. T. Grisaru, M. Rocek and W. Siegel,
  ``Superspace'', The Benjamin/Cummings Publishing Company, Reading (1983).}
\lref\GiNe{S. B. Giddings and W. M. Nelson, ``Quantum Emission from Two
  Dimensional Black Holes'', Phys. Rev. {\bf 46D}, 2486 (1992).}
\lref\RuTs{J. G. Russo and A. A. Tseytlin, ``Scalar Tensor Quantum
 Gravity in Two Dimensions'', Nuc. Phys. {\bf 382B}, 259 (1992).}

\Title{\vbox{\baselineskip12pt\hbox{UCSBTH-93-10}
\hbox{hepth@xxx/9304163}
}}
{\vbox{\centerline {$N=2$ Supersymmetry in}\vskip2pt
\centerline{Two-Dimensional
Dilaton Gravity}
}}

\centerline{{\ticp William M. Nelson}\footnote{$^\dagger$}
{Email address: nelson@denali.physics.ucsb.edu}
{\ticp and Youngchul Park}\footnote{$^*$}
{Email address: young@denali.physics.ucsb.edu}}
\vskip.1in
\centerline{\sl Department of Physics}
\centerline{\sl University of California}
\centerline{\sl Santa Barbara, CA 93106-9530}
\bigskip
\centerline{\bf Abstract}
Actions for $D=2$, $N=2$ supergravity coupled to a scalar field are
 calculated,
and it is shown that the most general power-counting renormalizable
dilaton gravity action has an $N=2$ locally supersymmetric extension.
The presence of  chiral terms in the action leads one to hope that
non-renormalization theorems similar to those in global SUSY will apply;
this would eliminate some of the renormalization ambiguities which
plague ordinary bosonic (and $N=1$) dilaton gravity. To investigate
this, the model is studied in superconformal gauge,
 where it is found
that one chiral term becomes nonchiral, so that only one term
is safe from renormalization.

\Date{4/93}

\newsec{Introduction}

The  theory of dilaton gravity, defined by the action
\eqn\DG{S_D = \int d^2 x \sqrt{-g} e^{-2 \phi}
              \bigl[ R + 4 (\nabla \phi)^2 + 4 \lambda^2 \bigr],
}
has received much attention recently as a two-dimensional model
of gravity  (see e.g. \refs{\CGHS,\RV});
as such it has proved to be a convenient
laboratory for studying many outstanding issues in black hole physics
and quantum gravity.
The theory, which derives originally from string theory,
 shares many interesting properties with four dimensional
general relativity,
$$ S_G = \int d^4 x \sqrt{-g} R.
$$
For example,
it has collapsing black hole solutions which radiate by the Hawking
process, and
 it has a positive energy theorem \refs{\PaSt}
analogous to the one for four dimensional general relativity.
In addition the theory is significantly simpler to work with than
general relativity, allowing many analytic results to be obtained
whose four-dimensional counterparts are known approximately at best.
In particular, the theory coupled to conformal matter is completely
soluble classically, and among the exact solutions one finds a set
describing collapse of matter to form a black hole. The black holes
emit Hawking radiation, for which the energy flux and Bogoljubov
coefficients
may be computed analytically; and, in addition, the
backreaction may be completely accounted for by a simple addition
to the action \refs{\CGHS,\GiNe} .

Unfortunately, this toy model is so faithful in resembling the four
dimensional general relativity, that it also shares some unwanted features
with the full theory; in particular,
although the theory is power-counting renormalizable, the fields are
dimensionless, so an infinite number of counterterms are
allowed, leading to  a quantum theory which is in principle the
most general power-counting renormalizable theory coupling $2D$ gravity
with a scalar dilaton. This large amibiguity
destroys predictability \refs{\Strm, \GiSt}.
One might expect that adding supersymmetry would ameliorate this
problem, since in global supersymmetry certain terms (chiral
terms) are known not to receive renormalization corrections.
$N=1$ supersymmetry (as explored in \refs{\PaSt}\ ),
although interesting in many ways,
 is not
sufficient for this purpose, since no chiral terms may be written down.
To have chiral superfields in $D=2$ one needs at least $N=2$
supersymmetry; accordingly,
in this paper, we  begin with
the most general power-counting renormalizable theory
of dilaton gravity
$$
S = \int d^2 x \sqrt{-g} \bigl[ J(\phi) R + 2 K(\phi) (\nabla \phi)^2
               + W(\phi) \bigr]
$$
and extend it to have $N=2$ supersymmetry.

\newsec{Formalism}

In the next section we will write actions using formulas first
calculated
in \refs{\Aln};
 however, since that paper contains errors, we reproduce
here the results we need. Some of these formulas are also given in
\refs{\GLO,\Howe}, and the method is that of \refs{\GGRS}, section 5.6.
 We use $N=2$ superspace with covariant
derivative
$$
\nabla_{A}={E_A}^{M} \partial_{M} + \omega_{A}\lor + \Gamma_{A}\why
$$
where  $\omega$ is the spin connection and $\Gamma$ is the $U(1)$
connection which arises due to $N=2$. Also $\lor$, $\why$ are the
Lorentz and $U(1)$ generators which act as follows on spinors
$F_{\alpha}$ and vectors $F_{a}$ :

$$ \eqalign{
[ \lor ,F_{\alpha} ]
      &= \hf {(\gamma^3)_{\alpha}}^\beta F_{\beta}, \cr
[\lor ,F_{a} ] &= {\epsilon_{a}}^{b} F_{b}, \cr
[\why ,F_{\alpha} ] &=
      \hf\imath {(\gamma^3)_{\alpha}}^\beta F_{\beta}. \cr }
$$
The gamma matrix conventions are
 ${(\gamma^0)_\alpha}^\beta = {(\sigma^2)_\alpha}^\beta$,
${(\gamma^1)_\alpha}^\beta = {( - \imath \sigma^1)_\alpha}^\beta$,
\hbox{${(\gamma^3)_\alpha}^\beta = {(\sigma^3)_\alpha}^\beta$,  }
and the metric is taken to have signature
$(+1, -1)$.
 Spinors are raised from the left
and lowered from the right by the spinor metric
$C_{\alpha\beta}=-C^{\alpha\beta}=\sigma^{2}$, and the antisymmetric
tensor $\epsilon_{ab}$ satisfies $\epsilon_{01}=1$.
With these conventions one has the identity
\hbox{${(\gamma^a)_\alpha}^{\dot \alpha}{(\gamma_a)_\beta}^{\dot \beta}
 =-{\delta_\alpha}^{\dot \alpha} {\delta_\beta}^{\dot \beta}
  - {(\gamma^3)_\alpha}^{\dot \alpha} {(\gamma^3)_\beta}^{\dot \beta}$.}
Conventions for conjugation are as in \refs{\GGRS}; in particular,
$(\psi^\alpha)^\dagger ={\bar\psi}^{\dot\alpha}$, while there is a
sign change if the indices are lowered, since the spinor metric
is imaginary. Lastly, we use the standard index convention whereby
indices from early in the alphabet are Lorentz indices while those
from later are Einstein.

To reduce to a minimal theory of supergravity, one imposes constraints on
the graded commutator $\left[\nabla_{A},\nabla_{B} \right\}$,
 after which one finds that the Bianchi identities may
be ``solved'' to express all the commutations in terms of one
chiral superfield, giving
 \refs{\Aln, \GLO ,\Howe}
\eqn\cnst{\eqalign{
\{\nabla_\alpha, \nabla_\beta \} &= 2 (\gamma^3)_{\alpha \beta} {\bar \R}
      ({\lor +\imath\why}), \cr
\{\nabla_{\alpha}, \gradbar_{\dot\alpha}\} &=
      2 \imath (\gamma^a)_{\alpha {\dot \alpha}} \nabla_a, \cr
[\nabla_\alpha, \nabla_a ] &=
      \hf \imath {(\gamma_a)_\alpha}^{\dot \alpha} [
      {\bar \R} \gradbar_{\dot \alpha} -
      {(\gamma^3)_\alpha}^{\dot \beta} \gradbar_{\dot \beta} {\bar \R} (
      \lor +\imath\why ) ], \cr
[\nabla_{a}, \nabla_{b} ] &= {1\over 4}
      \epsilon_{a b} [(\gamma^3)^{\alpha \beta}
      (\nabla_{\alpha} \R) \nabla_{\beta}
      +(\gamma^3)^{{\dot\alpha} {\dot\beta}} (\gradbar_{\dot\alpha}{\bar \R})
      \gradbar_{\dot\beta} ] \cr
      &+ {1\over 8}\epsilon_{ab} [ \nabla^{2} \R+\gradbar^{2}{\bar \R}
      -8\R{\bar \R} ]\lor
      -{\imath\over 8}\epsilon_{ab} [ \nabla^{2}\R -\gradbar^{2}{\bar \R}
      ] \why,  \cr }}
where $\R$ is a chiral superfield, $\gradbar_{\dot\alpha}\R=0$.

Supergravity component fields are defined as
$\theta ={\bar\theta}=0$ components of ${E_{a}}^M$, $\omega_a$, and
$\Gamma_a$ by
\eqn\sugracomps{ \eqalign{
\nabla_{a}| &= {e_{a}}^{m} \dee_{m} +
           {\psi_{a}}^{\mu} \dee_{\mu}
           +{{\bar\psi}_{a}}^{\dot\mu} \dee_{\dot\mu}
           +\omega_{a}\lor +A_{a}\why \cr
        &= {e_{a}}^{m} D_{m} + {\psi_{a}}^{\mu} \dee_{\mu}
           + {{\bar\psi}_{a}}^{\dot\mu} \dee_{\dot\mu}, \cr }
}
where ``$|$'' stands for the $\theta={\bar\theta}=0$ projection.
Then computing in a Wess-Zumino gauge, one finds
  the component content of $\R$  to be
\eqn\Rcomps{ \eqalign{
\R| &= B, \cr
\nabla_{\alpha}\R| &= -2\epsilon^{ab}
    {(\gamma^{3})_{\alpha}}^{\beta}\psi_{ab\beta}
    -2\imath B{(\gamma^b)_{\alpha}}^{\dot\beta}{\bar\psi}_{b\dot\beta}, \cr
\nabla^{2}\R| &= -2\epsilon^{ab} [r_{ab} +\imath f_{ab} ]
     +4B{\bar B} +16\imath {\bar\psi}^{a}\gamma^{b}\psi_{ab}
     +8{\bar\psi}^{a}{\bar\psi}_{a}B, \cr }
}
where $r_{ab}$,$f_{ab}$ and $\psi^{\alpha}_{ab}$ are respectively the
curvature, the $U(1)$ field strength, and the gravitino curl:
$$\eqalign{
r_{ab} &= {e_{a}}^{m} {e_{b}}^{n}\dee_{[m} \omega_{n]}, \cr
f_{ab} &= {e_{a}}^{m} {e_{b}}^{n}\dee_{[m} A_{n]}, \cr
\psi^{\alpha}_{ab} &= {e_{a}}^{m} {e_{b}}^{n}
              {D_{[m} \psi_{n]}}^\alpha. \cr }
$$
Our index suppression convention for  spinor indices is that
they are in matrix multiplication order, e.g.
$\, \bar\psi\gamma\psi\equiv
{\bar\psi}^{\dot\alpha}{\gamma_{\dot\alpha}}^{\alpha}\psi_{\alpha}$.

We will want to write down chiral terms in the action, which have the
generic form
\eqn\genericchiralterm{
\act = \int d^{2}\Theta\chden \act_{\rm chiral}. }
In this formula, $\act_{\rm chiral}$ is some chiral superfield,
$\gradbar_{\dot\alpha}\act_{\rm chiral} =0$. The integration is
over the two $\Theta$ variables of a chiral representation of
superspace; these are variables defined such that a chiral
superfield $\Phi$ has the $\Theta$ expansion
\eqn\chiralsfexpansion{
         \Phi=\Phi | +\Theta^{\alpha}\nabla_{\alpha}\Phi |
  -\frth{\Theta^2}\nabla^{2}\Phi |. }
One arrives at such a representation through a superspace
coordinate transformation whose parameters must depend on the fields
in the supergravity multiplet (since $\nabla_{\alpha}$ does).
To make the integral covariant under coordinate transformations
one needs the chiral density $\chden$; it serves the same
purpose as the ordinary spacetime density $\sqrt {-g}$.
To calculate $\chden$ there are various strategies, two of which
are given in \refs{\Aln,\WB}. With either, one finds
\eqn\chiraldens{ {1\over e}\chden    =
1 +2\imath\Theta\gamma^{a}{\bar\psi}_{a}
-\frth\Theta^{2}[-4{\bar
B}-8\epsilon^{ab}{\bar\psi}_{a}\gamma^{3}{\bar\psi}_{b}
]}
where $e=\sqrt{-g}$, and the factor of $-\frth$ is for convenience since
$\int d^{2}\Theta \Theta^{2} =-4$.
Lastly, chiral actions may be obtained from nonchiral with the projection
$$\act_{\rm chiral} = \gradbar^{2} \act_{\rm nonchiral}.$$

\newsec{Dilaton Gravity Action}

Now we would like to use this technology to
write actions which extend the bosonic and
$N=1$ dilaton gravity models to models with $N=2$, for the reasons stated
in the Introduction.
Since renormalization of dilaton gravity may produce in principle
any power-counting renormalizable term coupling a scalar dilaton to
two-dimensional gravity, we attempt to write down the most general
power-counting renormalizable action coupling a scalar dilaton to
two-dimensional $N=2$ supergravity. For $N=2$ supersymmetry the dilaton
must become a complex scalar in order to equalize bosonic and fermionic degrees
of freedom.
One minimal multiplet for a complex scalar field is the chiral scalar
multiplet, consisting of the components of one chiral superfield, which
we call $\Phi$.
We find  three renormalizable terms can be written, corresponding to the
three terms in the original dilaton gravity action \DG.
\eqn\action{
\act = \int d^2 \Theta \chden
       \bigl[ \R S(\Phi) +
          \gradbar^2 T(\Phi,\bar\Phi) +
          U(\Phi) \bigr] + \hbox{   h.c.}
}
The functions $S$, $T$, and $U$ are arbitrary.

In order to see what models of dilaton gravity are generated by this
theory, we study its classical bosonic solutions; for this purpose
we need only the bosonic part of the action, which comes out to be
$$\eqalign{
{1\over e} \act &= -4R\Re S(\pphi)
                   +4F\Im S(\pphi)
        +32\Re\left[ \deebar\dee T(\pphi, {\bar \pphi}) \right]
         \nabla^{a}{\bar \pphi} \nabla_{a}\pphi \cr
        &+ 2\Re\left[ U^{\prime}(\pphi)G-4 U(\pphi){\bar B}
          + S^{\prime}(\pphi)BG\right] \cr
           &+ 2\Re\left[ \deebar\dee T(\pphi, {\bar \pphi})\right]
               G{\bar G}, \cr }
$$
where
$$
\eqalign{
\Phi| &= \pphi , \cr
{\nabla^2} \Phi| &= G. \cr}
$$
and
$ R = \epsilon^{ab}r_{ab} $ is the scalar curvature, while
$F \equiv\epsilon^{ab}f_{ab}$ is the equivalent quantity for the $U(1)$
field.

Next we eliminate the auxiliary fields from the  action, beginning with
 $B$. Variation of $B$   leads to the equation
$$G= {4{\bar U}\over S^{\prime}};$$
when this is substituted back into the action, all terms with $B$
cancel - as expected, since $B$ appeared only linearly.
Variation of $G$ leads to an equation for $B$, which is irrelevant,
since $B$ has been eliminated.

Next we consider the equation of motion of the $U(1)$ field $A_{m}$.
This field appears only once in the bosonic action, in the field
strength $F$. Its variation yields
$$\partial_{m}\Im [S(\pphi)]=0,$$
i.e., $\Im S ={\rm const.}\equiv c$.
This we solve parametrically, introducing a real parameter $\phi$ and
setting $\pphi=\pphi(\phi)$ such that $\Im S(\pphi(\phi))=c$.
Then for functions $F(\pphi)$ we have
$$F(\pphi)\rightarrow F(\phi)\equiv F(\pphi(\phi)) $$
and
$$F^{\prime}(\pphi)\rightarrow F^{\prime}(\phi)\left({d\pphi\over
    d\phi}\right)^{-1} .$$

Lastly, we must satisfy  equation of motion of the degree of freedom
which was eliminated with $\pphi=\pphi(\phi)$; this is equivalent
to varying $c$, which just fixes $F=0$.

Making these substitutions in the action,
and redefining
\eqn\redfn{
\eqalign{
J(\phi) &= 4 S(\phi), \cr
K(\phi) &= - 16\left|{d\pphi\over d\phi}\right|^{2}
                  \Re [\deebar \dee T(\phi, \phi)], \cr
L(\phi) &= 8 U(\phi),\cr}
}
one finds that  $c$ disappears, so that $J$ may be taken to be real.
Furthermore, $L$ appears only in the forms
 $|L|^2$ or $\Re (L^{\prime}\bar L)$,
which are unchanged if $L\rightarrow |L|$; therefore we may take $L$
to be real. The action we finally obtain is
\eqn\fnact{
{1\over e} \act = -RJ - 2 K (\nabla\phi)^{2}
     + {{L L^{\prime}} \over {2 J^{\prime}}}
     - {{K L^{2}} \over 2 {( J^{\prime})^{2}}},
}
which is almost exactly the action obtained in the $N=1$ case
 \refs{\PaSt}; the only difference is the sign of the first two terms,
which is just due to our opposite-signature
metric.
The stationary points of this action are in one to one
correspondence with the
bosonic classical solutions of the full action \action.
It is curious that the bosonic solutions of this $N=2$ theory are
 the same as those of the $N=1$ theory, even though here there
is initially an extra bosonic degree of freedom.

\newsec{Superconformal Gauge}

In this section we study the theory \action\ in superconformal
gauge in superspace, both because it illuminates the question of
non-renormalization of chiral terms, which we wish to
understand, and because it is interesting in its own right.

The action \action\ as it stands is not ready for quantization,
since no gauge has been chosen and no superfields have been
identified for expansion (curved space chiral superfields such as $\Phi$
above are not suitable for quantization, since the chirality condition
is a constraint which depends on the supergravity fields.)
A good gauge for quantizing bosonic
dilaton gravity is  conformal gauge (see e.g. \refs{\GiSt}),
$g_{ab}=e^{2\rho}\eta_{ab}$; this completely fixes the gauge freedom,
and the fields $\phi$ and $\rho$ are the quantum fields (along
with some ghosts). One expands the action in $\phi$
and $\rho$ about the free
action and quantizes as usual. In the present case we adopt the
analogous solution, choosing a superconformal gauge
in superspace, which was
shown to exist  in \refs{\Howe}.
This gauge corresponds to the component field conditions
$$\eqalign{ {e_{a}}^{m} &= e^{-\rho}{\delta_{a}}^{m}, \cr
            \psi^{a} &= \gamma^{a}\chi, \cr
            \partial^{m} A_{m} &= 0, \cr }
 $$
where $\chi$ is a Dirac spinor .
Note that the last equation is the integrability condition for the existence
of $\alpha$ satisfying $A_{m}={\epsilon_{m}}^{n}\partial_{n}\alpha$. The
fields $\rho$, $\chi$, and $\alpha$, plus the complex auxiliary field $B$,
have altogether $4+4$ real degrees of freedom, suggesting that they might
fit into one chiral superfield. So one  guesses that
in this theory superconformal gauge is realized by a chiral scalar
field $\Lambda$, which is the ``superconformal factor''
extending the ordinary conformal factor $\rho$. Alternatively, one
can observe as in \refs{\Howe}\ that the Bianchi identities were
solved in terms of a single chiral superfield $\R$, which must
therefore contain all the gauge invariant information in the theory.

The form of the super-Weyl transformations in $N=2$ superspace was
worked out  in \refs{\Howe}; in terms of the quantities
${H_{A}}^{B}\equiv {E_{A}}^{M}\delta {E_{M}}^{B} $ they are
\eqn\superweyl{
       \eqalign{ {H_{a}}^{b}&=
(\delta\Lambda +{\delta\bar\Lambda}){\delta_{a}}^{b}, \cr
         {H_{\alpha}}^{\beta}&=
{\delta\bar\Lambda}{\delta_{\alpha}}^{\beta}, \cr
 	{H_{a}}^{\alpha} &=
\imath(\gamma_{a})^{\alpha\beta}D_{\beta}\delta\Lambda. \cr }
}
Here $D_{\alpha}$ is the usual flat superspace covariant derivative
$D_{\alpha} = \partial_{\alpha}+\imath{\bar\theta}^{\dot\alpha}
   (\gamma^{a})_{\alpha\dot\alpha}\partial_{a} $.

Already from here one can derive the interesting result that
$E\equiv {\rm sdet}({E_{M}}^{A}) =1$
in conformal gauge; i.e., the superdeterminant
of the vielbein is pure gauge.  This follows  from
$\delta E = E {\rm str}(E^{-1}\delta E) = E {\rm str}H=0$
(see e.g. \refs{\GGRS}, section 3.7).
So the chiral action $\int d^{2}\Theta \chden \R $ cannot be written as
a full superspace integral $\int d^2\theta d^2{\bar\theta} E$;
in this respect the theory differs from $D=4$, $N=1$ supergravity.
This leads one to hope that the $S$ term in
the action \action\ will in fact
be protected from renormalization. Unfortunately, the conformal gauge
analysis doesn't bear this out.

Working from the variations \superweyl\ , one can construct the
superconformal gauge forms of all superfields in the theory.
We give the ones we need in the form
\eqn\scgforms{
 \eqalign{
  \nabla_{\alpha} &= e^{-\bar\Lambda}D_{\alpha}
     -e^{-\bar\Lambda}{(\gamma^3)_{\alpha}}^{\beta}D_{\beta}\Lambda
           (\lor+i\why), \cr
  \gradbar_{\dot\alpha} &= e^{-\Lambda}{\bar D}_{\dot\alpha}
     -e^{-\Lambda}{(\gamma^3)_{\dot\alpha}}^{3\dot\beta}
     {\bar D}_{\dot\beta}{\bar\Lambda}
           (\lor-i\why), \cr
  \R &= -\hf e^{-2\Lambda}{\bar D}^{2}{\bar\Lambda}. \cr
} }
\noindent From here one finds $\nabla_{a}$ using the constraint on
$\{\nabla_{\alpha},\gradbar_{\dot\alpha}\}$ (eq. \cnst\ );
it comes out to
\eqn\nablaa{\nabla_{a}=e^{-\Lambda-{\bar\Lambda}}\left[
 \partial_a -{i\over 2}(\gamma_a)^{\beta\dot\beta}\left(
 D_{\beta}\Lambda{\bar D}_{\dot\beta}
     +{\bar D}_{\dot\beta}{\bar\Lambda}D_{\beta} \right)\right]
+{\rm connections}, }
and one can go on to
  deduce the vielbein, inverse vielbein, etc.
These equations can be  checked by verifying they satisfy the
constraints \cnst.

A useful observation is that the form of $\nabla_\alpha$ given above,
along with
 $ [(\lor+i\why),\psi^{\alpha}]=0$,
 implies
$$\nabla^{2}P=e^{-2 \bar \Lambda} D^{2}P ,$$
where $P$ stands for
any scalar superfield; and analogously for $\gradbar^{2}P$.

For our purposes,
it is convenient to define the component content of $\Lambda$ by
comparing the expressions for $\R$ in \scgforms\
and in \Rcomps\ ;
this gives
\eqn\lambdacomp{ \eqalign{
 \Lambda | &= \lambda \equiv {1 \over 2} (\rho -i \alpha), \cr
 D_{\alpha}\Lambda | &= ie^{\bar\lambda}{(\gamma^b)_\alpha}^{\dot\alpha}
           {\bar\psi}_{b\dot\alpha}, \cr
 D^{2}\Lambda | &=-2{\bar B} e^{2\bar\lambda} .\cr
} }
(Note that this is not what one finds by comparing \nablaa\ and
\sugracomps\ ; the reason is that the computations of section $2$
were in  WZ gauge, which is not compatible with superconformal gauge.)

In order to write the action \action\ in conformal gauge,
we need the form of $\chden$. The simplest guess is $e^{2\Lambda}$,
but this doesn't satisfy $\chden | =e$. The correct form is
$$\eqalign{\chden &= Fe^{2\Lambda}\cr
       {\rm with} \,\,\,     F &= \exp\left[ 2{\bar\lambda}\left(
              x^\mu+i{\bar\theta}{\gamma^\mu}\theta \right) \right].\cr
 } $$
The variable substitution in $F$ is the familiar transformation leading
to the chiral representation in global supersymmetry. It gives $F$ the
convenient properties
$$\eqalign{ F| &= e^{2\bar\lambda}, \cr
            {\bar D}_{\dot\alpha} F &=0, \cr
          D_\alpha F| &= D^2 F|=0.\cr}
$$
 This formula for $\chden$
can be checked explicitly using the component forms
\chiraldens\ and \lambdacomp.

Finally we note using \chiralsfexpansion\ that
\genericchiralterm\ may be written in the representation
independent form
\eqn\repind{ \act = \nabla^{2}\chden\act_{\rm chiral} |, }
which can be immediately translated into conformal gauge using the
above formulas. Applying this procedure to the dilaton gravity action
\action\ yields
\eqn\cgaction{\eqalign{
  \act &= -\hf D^{2}{\bar D}^{2}({\bar\Lambda} S) |
          +D^{2}{\bar D}^{2} T | +D^{2}e^{2\Lambda} U | +{\rm h.c.}\cr
       &=\int d^{2}\theta d^{2}{\bar\theta}
        (-\hf {\bar\Lambda}S-\hf\Lambda{\bar S} + T+{\bar T})
         +\int d^{2}\theta e^{2\Lambda}U
         +\int d^{2}{\bar \theta} e^{2 \bar \Lambda} {\bar U}. \cr
} }
Note that in computing the component action from this
and comparing with \fnact\ one must remember
that components of $\Phi$ were defined previously by
$\nabla_{\alpha}\Phi |$, etc., and are related by factors of
$e^{\lambda}$, $e^{\bar\lambda}$ to components defined with
$D_{\alpha}$. To this action must be added of course
Fadeev-Popov ghosts; we do not compute those terms here.

An unfortunate feature of this action is that the $S$ term is now a
full superspace integral, i.e. it is not protected from
renormalizations as we had hoped. Only the $U$ term remains chiral,
and should remain unaffected by renormalizations.

Note that we cannot get \cgaction\ by simply
$N=2$ supersymmetrizing the
bosonic $\sigma$-model action,
\eqn\bosact{
\act = \int d^2 x [ 8 \partial_+ \partial_- \rho J(\phi)
            - 8 K(\phi) \partial_+ \phi \partial_- \phi
            + e^{2 \rho} W(\phi)]
}
because the target space $(\rho, \phi)$ does not have a metric with
definite sign and we need a K\"ahler target space
to $N=2$ supersymmetrize and  K\"ahler metrics have
definite sign.

\newsec{Discussion}

As  mentioned in the Introduction,
quantization of the dilaton gravity action \fnact\
is ambiguous because of the infinite number of  allowed
counter-terms. Our attempted cure is to extend the
action to have $N=2$ supersymmetry where we can write chiral
action terms which we hope won't suffer renormalizations.
Unfortunately this program meets with only partial success, since
only one term remains chiral when we put the theory in conformal
gauge for quantization purposes. But for that term the
the global supersymmetry non-renormalization theorem
should apply.

An interesting feature of the bosonic action \bosact\ is that
classically it really only
contains one arbitrary function; $J$ may be varied at will,
and $K$ may be eliminated,
by appropriate (non-derivative)
redefinitions of the fields $\phi$ and $\rho$ \refs{\RuTs}.
One wonders whether this
operation can be extended  supersymmetrically to the whole theory;
from the superconformal gauge action \cgaction\ one
sees that it cannot, since a shift
$\Lambda\rightarrow\Lambda + Y(\Phi)$ causes a change
\hbox{$T+{\bar T} \rightarrow T+{\bar T} -\hf {\bar Y}S-\hf Y{\bar S}$,}
and it is not possible in general to cancel $T$ in this way.
 The obstacle is the chirality of $\Lambda$ and
$\Phi$ (which must be preserved so that the $U$ term remains
supersymmetric); this problem is absent in the $N=1$ theory
of \refs{\PaSt}, so in that theory the field redefinitions should
be possible. Furthermore due to supersymmetry there is an extra
bonus, namely that such non-derivative transformations produce
trivial Jacobian in the functional integral
(\refs{\GGRS}, section 3.8); however, they will alter the
measures for the $\Phi$ and ghost path integrals, and will also alter
any additional matter action which is coupled to the theory.
So their significance is unclear.

Finally we note that in $D=2$, there is another $N=2$  superspace
geometry \refs{\GLO, \Howe}, in which the $U(1)$ action on spinors involves
${\delta_{\alpha}}^{\beta}$ instead of ${(\gamma^3)_{\alpha}}^{\beta}$,
 and another scalar multiplet (the twisted
chiral multiplet). We could have attempted to use a different combination
of these than the one we chose;
 however, it appears that only the fermionic sector would
be changed.

Possible directions which are not explored in this paper include
further pursuing the quantization of these models,
 and extending the supersymmetry to $N=4$.

\bigskip\bigskip
\centerline{{\bf Acknowledgments}}\nobreak
We would like to thank
  M. Rocek and P. Townsend, for useful
conversations;  S. J. Gates, for a helpful e-mail
comment; and S. Giddings and A. Strominger for discussions and
comments on earlier drafts.  This work was supported in
part by the grants DOE-91ER40618 and NSF PYI-9157463.

\listrefs

\end